\documentclass[aps,preprint,showpacs,preprintnumbers,amsmath,amssymb]{revtex4}
\usepackage{amsmath,mathrsfs,amsbsy,color,graphicx,bm,amsthm,amsfonts}
\usepackage{units}
\usepackage{bbm}
\usepackage{times}
\usepackage{dcolumn}
\usepackage{mathrsfs}
\usepackage{amsmath,amssymb,epsfig,float}

\begin{document}

\title{ Multipartite quantum clock synchronization under the influence of Unruh thermal noise}

\author{Li Zhang}
\affiliation{Department of Physics, Collaborative Innovation Center for Quantum Effects, and Key Laboratory of Low
Dimensional Quantum Structuresand Quantum \\
Control of Ministry of Education,
 Hunan Normal University, Changsha, Hunan 410081, China}

\author{Jiliang Jing}
\affiliation{Department of Physics, Collaborative Innovation Center for Quantum Effects, and Key Laboratory of Low
Dimensional Quantum Structuresand Quantum \\
Control of Ministry of Education,
 Hunan Normal University, Changsha, Hunan 410081, China}

\author{Heng Fan}
\affiliation{Beijing National Laboratory for Condensed Matter Physics, Institute of Physics,Chinese Academy of Sciences, Beijing 100190, China}

\author{Jieci Wang}
\email{jcwang@hunnu.edu.cn}
\affiliation{Department of Physics, Collaborative Innovation Center for Quantum Effects, and Key Laboratory of Low
Dimensional Quantum Structuresand Quantum \\
Control of Ministry of Education,
 Hunan Normal University, Changsha, Hunan 410081, China}

\begin{abstract}

\vspace*{0.2cm} We perform a protocol for multipartite quantum clock synchronization under the influence of Unruh thermal noise. The clocks consisting of  Unruh-DeWitt detectors when one of  detectors accelerated is obtained.  To estimate the time difference between the clocks, we calculate the  time probability and analyze how the probability is influenced by the Unruh thermal noise and other factors. It is shown that both relativistic motion and interaction between the atom and the external scalar field  affect the choice of optimal number of  excited  atoms in the initial state, which leads to a higher clock adjustment accuracy.  Time probabilities for different types of initial states approach to the same value  in the limit of infinite acceleration, while  tend to  different minimums with increasing number of atoms.  In addition, the accuracy of clock synchronization using a pair of entangled clocks in two-party  system
is always higher than   in an multipartite system, while the  optimal $Z$-type multipartite initial state always  perform better than the $W$-type state.

\end{abstract}

\maketitle

\section{Introduction}

High-accuracy synchronization of separated  clocks  is of fundamental interest in  fundamental physics and  a wide range of applications such as the global positioning system (GPS),  navigation, gravitational wave observation,  as well as the laser interferometer gravitational wave observation (LIGO) \cite{Simons,Lewandowski}. There are two approaches to  synchronize  a pair of clocks. Einstein's synchronization proposal \cite{Einstein} is implemented by  exchanging pulses between two spatially separated clocks and measuring their arrival time. An alternative approach is the  Eddington clock transport proposal \cite{Eddington}, which is based on sending a locally synchronized clock from one part to the other part.  However, both of them are limited by the unstable intervening media, and a few quantum strategies of clock synchronization \cite{qsyn41,qsyn5,qsyn6,yuefan1,yuefan2,qsyn71,qsyn72,Ren,Komar} and  experiments \cite{Valencia,Quan,Zhang} have been performed to achieve an enhancement of accuracy. In these studies,  precisions of clock synchronization in bipartite \cite{Jozsa,Yurtserver,Burt,Jozsa2} and multipartite\cite{qsyn3,qsyn4,Krco,Ben-Av}  quantum systems were found to be greatly improved by quantum resources.

On the other hand, the feasibility of QCS under the influence of clocks' relativistic motion has been
paid much attention because time dilation induced by relativistic effects  is experimentally observed due to velocities of several meters per second by comparing a  moving clock with a stationary clock \cite{Alclock}. 
We studied the influence of relativistic  gravitational frequency shift effect on satellite-based QCS and found that the precision of QCS is remarkably affected by the relativistic effect of the Earth \cite{wangsyn}.
When it comes to the accelerated clock, the Unruh effect \cite{Unruh,Crispino},  a significant prediction in quantum field theory is necessary to be considered. The Unruh effect shows that a uniformly accelerated observer would see a thermal bath, whose temperature is proportional to the proper acceleration possessed by observers. The influence of Unruh effect on quantum systems \cite{RQI0,RQI01,RQI1,RQI2,RQI3,RQI31,RQI7,RQI4,RQI5,RQI6} is a focus of study recently because such studies provide insights into some key physical questions such as nonlocality, causality, and the information paradox of black holes. Since it's hard to observe the Unruh effect in laboratory, people approache to the Unruh's original derivation with the semiclassical Unruh-DeWitt detector model \cite{UW84}.  Recently, dynamics of  entanglement \cite{Landulfo}, quantum discord \cite{Landulfo1},  quantum nonlocality \cite{Tian}, and quantum coherence \cite{Wang3} of two entangled detectors under the influence of  Unruh thermal noise have been discussed.  Although some  efforts have been devoted to study the influence of Unruh effect on quantum information,  these studies are confined to two-detector systems, while the behavior of many-detector system in an accelerated frame is an ongoing topic of research.  

In this paper, we perform a multipartite protocol for QCS in a relativistic frame, where each clock corresponds to an two-level atom modeling  by the Unruh-DeWitt detector.
We assume that $n$ detectors are prepared in an entangled initial state. The difference between the proper time of inertial reference frames and accelerated reference frames is determined by the observable time probability of  final state.  Unlike previous multipartite QCS protocols \cite{Krco,Ben-Av}, we concern about the  QCS for an accelerated clock and the influence of Unruh thermal noise is involved.  Without loss of generality, we assume that the first static clock in the $n$-party system has been  synchronized and  regarded as the standard clock.
There are three other advantages of  multipartite systems that motivate us.
Firstly,  we can synchronize more clocks with only one standard clock  in many-clock systems than in two-clock systems. This public, classical information  allows all parties to synchronize their time read to the standard clock. That is to say,  public results are accessible at the convenience of each party.
Secondly, many-detector systems provide more approaches to read measurement data than two-detector systems. Expect accelerated clocks, other clocks remain static and  could be the standard clock.
Hence, we could also obtain the standard time information from other clocks that might be a standard clock if necessary.
Thirdly, maximal entanglement between pairs of qubits is not necessary for multipartite clock synchronization \cite{Krco}. Because the monogamy of entanglement for  the multipartite system, no two of these qubits should be maximally entangled.
And multipartite entanglement has become a valuable resource in general relativity, quantum mechanics and quantum information, including quantum computation and quantum communication\cite{DGJB,LPMG,DBCM,SLHS}. It is impressive that a multipartite entangled state with 20 controlled particles  has been generated experimentally\cite{NFOM}.

The outline of this paper is organized as follows. In Section.~\ref{sec:relativistic}, we study the evolution of the multipartite system when one clock is carried by the accelerated detector. In Sec.~\ref{sec:Msynchronization}, we study the QCS for an accelerated clock in the multipartite system  and analyze how the Unruh thermal noise and other factors affect the accuracy of  our scheme.   In Sec.~\ref{sec:conclusion}, we make a conclusion.

\section{ Evolution of the multipartite relativistic quantum system }
\label{sec:relativistic}

To synchronize $n$ spatially separated clocks, Krco and Paul \cite{Krco}  promoted a multipartite synchronization protocol, in which any one of the clocks can be taken as the standard clock. In their proposal, the initial shared among these clocks is a symmetric W-type state \cite{Dur}
\begin{eqnarray}\label{Multiple1}
|\psi_{W}\rangle=\frac{1}{\sqrt{n}}(\overbrace{|10...00}^{n\;\;atoms}\rangle+|01...00\rangle+...+|00...01\rangle),
\end{eqnarray}
which involves $n$ subsystems and only one of subsystems lie in the excited state.  A few years later, Ben-Av and Exman employed the $Z$-type state as a generalization of $W$ state   and found that the clock adjustment accuracy was improved via $Z$-type state properties \cite{Ben-Av}. To study the dynamic of  many-detector system under relativistic motion, we are going to work out the evolution of  $Z$-type initial state in a relativistic setting firstly. The dynamic of $W$-type state can be obtained in a similar way.   The $Z$-type state is an entangled state with $n$ two-level noninteracting atoms and fully symmetric under the operation of particle exchange, which owns the form
\begin{eqnarray}\label{Multiple1}
 |\psi_{t_{0}}\rangle
 & = & \sqrt{\frac{(n-k)!k!}{n!}}(|111...000\rangle
 + |11...01...00\rangle+...+|000...111\rangle),
\end{eqnarray}
where $k$ atoms are involved in the excited energy eigenstate $|1\rangle$, while the rest lie in the ground state $|0\rangle$. In our scheme of clock synchronization, each atom must be an energy eigenstate  with  with a well-defined energy so that the density matrix of initial state is constant and  known to us before measurements. And the initial state does not have to be symmetric. The symmetry of $Z$-type multipartite initial state which is a natural property of our systems is only just convenient to illustrate the process of clock synchronization. Besides, $Z$-type initial state is helpful for us to consider  the influence of  initial state for our QCS with different values of total atoms $n$ and excited atoms $k$.

The total Hamiltonian of whole system is
\begin{equation}\label{tothamitonian}
  H_{n\phi}=\sum_{i=1}^{n}H_{i}+H_{KG}+H_{int}^{R\phi},
\end{equation}
where $H_{KG}$ stands for the free Hamiltonian of  massless scalar field satisfying the Klein-Gordon equation $\Box\phi=0$.
$H_{i}=\Omega D_{i}^{\dagger} D_{i},i=1,2...n$ is the free Hamiltonian of each atom, where $D_{i}$ and $D_{i}^{\dagger}$ represent the creation and annihilation operators of the $i$th atom, respectively.  And $\Omega$ is the energy gap between the ground state $|0\rangle$ and the excited energy eigenstate $|1\rangle$.
In Eq. (\ref{tothamitonian}), $H_{int}^{R\phi}$ means interaction Hamiltonian between the atom and the massless scalar field.

We assume that the second atom of multipartite system, carried by an observer Bob, is uniformly accelerated for a duration time $\Delta$, while other atoms  keep static and are switched off all the time. The running of Bob's clock is affected by the relativistic motion and needs to be synchronized.
The world line of  Bob's detector is described by
\begin{equation}
 t(\tau_{B})=a^{-1}\sinh a \tau_{B}, x(\tau_{B})=a^{-1}\cosh a \tau_{B}, y(\tau_{B})=z(\tau_{B})=0,
\end{equation}
where  $a$ is the proper acceleration and $\tau_{B}$ is the proper time of Bob.
The interaction Hamiltonian $H_{int}^{B\phi}$ between Bob's detector and  massless scalar field is
\begin{equation}\label{hamitonian}
H^{R\phi}_{int}(t)=
\epsilon(t) \int_{\Sigma_t} d^3 {\bf x} \sqrt{-g} \phi(x) [\chi({\bf x})D_{2} +
                           \overline{\chi}({\bf x})D_{2}^{\dagger}],
\end{equation}
where $g_{ab}$ is the Minkowski metric and $g\equiv {\rm det} (g_{ab})$.
 Here the smooth compact-support real-valued function $\epsilon(t)$  keeps the detector switched on for a finite amount of proper time $\Delta$ (for more details on finite-time detectors see Ref.\cite{Higuchi}). After that, Bob turns off the accelerating engine  and comes to rest again. In Eq. (\ref{hamitonian}), $\Sigma$ is some suitable timelike isometries followed by Bob's accelerated detector, where coordinates $\bf{x}$ are defined and the spacetime point is donated by $x$.  For a given Minkowski time $t$, we could find corresponding  Klein-Gordon  field operators $\phi(x)$ on the static slice $\Sigma_{t}$. $\chi(\mathbf{x})$ vanishes outside a small volume around the accelerated atom. In that case, the integration   on the global spacelike Cauchy surface $\Sigma_{t}$  indicates that the interaction Hamiltonian in Eq.$(\ref{hamitonian})$ only appears  in the neighbourhood of Bob's detector.

By introducing a compact support complex function $ f(x)=\epsilon(t)e^{-i\Omega t}\chi({\bf x})$, we have $\phi(x)f\equiv Rf-Af$, where $A$ and $R$ are the advanced and retarded Green functions. Then we obtain
\begin{equation}\label{phif}
\phi(f)\equiv \int d^4 x \sqrt{-g}\phi(x)f=i[a_{RI}(\Gamma_{-}^{*})-a_{RI}^{\dagger}(\Gamma_{+})],
\end{equation}
where  $\Gamma_{-}$ and $\Gamma_{+}$  represent  negative and positive frequency parts of $\phi(f)$, respectively.  $a^{\dagger}_ {RI}$ and $a_{RI}$ are Rindler creation and annihilation operators in the right Rindler wedge $I$. Since $\epsilon(t)$ is a roughly constant for $\Delta\gg\Omega^{-1}$, the test function $f$ approximately owns the positive-frequency part, which means $\Gamma_{-}\approx0$. And if we define $\lambda\equiv -\Gamma_{+}$, Eq. (\ref{phif}) could be rewritten as $\phi(f) \approx i a^{\dagger}(\lambda)$.

Here we only consider the first order approximation under the weak-coupling limit. With Eq. (\ref{hamitonian}) and Eq. (\ref{phif}), the final state of  whole system in the interaction picture at time $t>t_{0}+\Delta$ is found to be

\begin{eqnarray}
|\Psi_{t}^{n\phi}\rangle
& = & |\Psi_{t_{0}}^{n\phi}\rangle +\frac{1}{(1-q)^{1/2}}\sqrt{\frac{(n-k)!k!}{n!}}
 ( |\Psi_{1}^{n\phi}\rangle\otimes q^{1/2}|1_{F_{2\Omega}}\rangle
 +  |\Psi_{2}^{n\phi}\rangle\otimes|1_{F_{1\Omega}}\rangle\rangle),
\label{Wholefistate}
\end{eqnarray}
where the parameterized acceleration $q \equiv e^{-2\pi\Omega / a}$ have been introduced. The whole initial state of $n$-party systems and the massless scalar field is $|\Psi_{t_{0}}^{n\phi}\rangle=|\psi_{t_{0}}^{n\phi}\rangle\otimes|0\rangle_{M}$, where $|0\rangle_{M}$ stands for Minkowski vacuum.
Bob's  creation and annihilation operators have been donated by $D_{2}^{\dagger}$ and $D_{2}$, respectively.
In Eq.(\ref{Wholefistate}),
\begin{equation}
|\Psi_{1}^{n}\rangle=D_{2}^{\dagger}|\Psi_{t_{0}}^{\phi}\rangle=(\underbrace{|11...00\rangle+...+|01...11\rangle}_{C_{n-1}^{k}\;\;terms}),
\end{equation}
where Bob's atom  and $\frac{(n-1)!}{(n-k-1)!k!}$ atoms of  the rest atoms are sure to lie in excited energy eigenstates $|1\rangle$.
\begin{equation}
|\Psi_{2}^{n}\rangle=D_{2}|\Psi_{t_{0}}^{\phi}\rangle=(\underbrace{|10...00\rangle+...+|00...11\rangle}_{C_{n-1}^{k-1}\;\;terms}),
\end{equation}
where Bob's atom  is certain to be in ground state $|0\rangle$
and $\frac{(n-1)!}{(n-k)!(k-1)!}$ atoms in  the rest atoms are sure to be involved in excited energy eigenstates.

Since we only concern about the final state of  multipartite system after Bob accelerated, the degrees of freedom of external scalar field should be  traced  out
\begin{equation}
  \rho_{t}^{n} = ||\Psi_{t}^{n\phi}||^{2} \textrm{tr}_{\phi}|\Psi_{t}^{n\phi}\rangle
  \langle\Psi_{t}^{n\phi}|.
\end{equation}
Using the fact that $(F_{i\Omega},F_{j\Omega})_{KG}=||\lambda||^{2}\delta_{ij}$ ($i\in{1,2}$),
we obtain
\begin{eqnarray}\label{Multiple2}
  \rho_{t}^{n}
  &=& |C|^{-2}\{|\psi_{t_{0}}\rangle\langle\psi_{t_{0}}|+\frac{n!}{(n-k)!k!}\frac{\nu^{2}}{1-q}
   (q|\Psi_{1}^{n}\rangle\langle\Psi_{1}^{n}| +|\Psi_{2}^{n}\rangle\langle\Psi_{2}^{n}|)\},
\end{eqnarray}
where
$\nu^{2}\equiv||\lambda||^{2}=\frac{\epsilon^{2}\Omega\Delta}{2\pi}%
e^{-\Omega^{2}\kappa^{2}}$ is  the effective coupling and $C=(1+\frac{q\nu^{2}(n-k)+\nu^{2}k}{(1-q)n})^{1/2}$ normalizes the final state $\rho_{t}^{n} $.

\section{QCS for the accelerated  clock in a  multiparty quantum system}
\label{sec:Msynchronization}
In this section, we study the QCS in  many-detector quantum systems. To synchronize Bob's accelerated clock, we select the first atom carried by a static observer, Alice, as the  standard clock. In the beginning,  Alice and Bob make an  appointment to  the starting  time $\tau$ to measure their local state. However, since Bob's clock has been accelerated, they experience different proper times. Therefore, Alice and Bob  can only start their measurements at their own proper times $\tau_A=\tau $ and $\tau_{B}=\tau$ respectively, which are relative to  time readings of their own
clock and are different due to relativistic effects. Once Bob have finished the accelerated motion and Alice's clock points at $\tau_{A}=\tau$, Alice measures her state immediately, which reduces the collapse of wave packet of the entangled state shared between Alice and Bob. Then Bob also measures his state at the  appointed proper time $\tau_{B}=\tau$, which is different from  Alice's measuring time due to Bob's relativistic motion. Then we obtain the time probability  which is used to estimate the   difference between Alice's and Bob's proper time.  It is worth to note that  in our proposal, the time difference information induced by Bob'€™s
relativistic motion is encoded in the probability $P$, and Bob is required to synchronize his clock  depending on what
he measures for P for a single clock. According to the existing clock synchronization proposals \cite{Jozsa,Krco},  local measurement and classical communication  would enable Bob to synchronize his clock to the standard clock as soon as the  time probability $P$ is obtained.

Since we only need to  synchronize Bob's clock according to Alice's standard clock, the degree of freedom for all the rest atoms should be traced out. Indeed, any accelerated clock in the $n$-party system can be synchronized with  Alice's clock, whereas any static clock can be selected as the standard clock.
Tracing out the the rest sub-sytems in Eq. (\ref{Multiple2}), we obtain the reduced density matrix for Alice and Bob \label{F1}
\begin{equation}\label{density1}
  \rho^{AB}= \frac{k(n-k)}{|c|^{2}n(n-1)}\begin{pmatrix}
                                    S_{1} & 0 & 0 & 0 \\
                                    0 & S_{2} & 1 & 0 \\
                                    0 & 1 & S_{3} & 0 \\
                                    0 & 0 & 0 & S_{4}
                                  \end{pmatrix},
\end{equation}
in their own basis ($|00\rangle,|01\rangle,|10\rangle,|11\rangle$), where $S_{1}=\frac{n-k-1}{k}+\frac{\nu^{2}}{1-q}$, $S_{2}=1+\frac{q\nu^{2}(n-k-1)}{(1-q)k}$, $S_{3}=1+\frac{\nu^{2}(k-1)}{(1-q)(n-k)}$ and $S_{4}=\frac{k-1}{n-k}+\frac{q\nu^{2}}{1-q}$.

To synchronize Bob's clock, we adopt the dual basis $|pos\rangle=\frac{1}{\sqrt{2}}(|0\rangle+|1\rangle)$ and $|neg\rangle=\frac{1}{\sqrt{2}}(|0\rangle-|1\rangle)$ as the measurement basis, which are obtained from $|0\rangle$ and $|1\rangle$ through the Hadamard transformation. In the measurement basis, Eq.  (\ref{density1}) is rewritten into
\begin{equation}
  \rho^{AB}=\frac{1}{8+4\alpha_{+}}\begin{pmatrix}
                                   4+\alpha_{+} & \beta_{+}&\alpha_{-} & -4+\beta_{-}  \\
                                    \beta_{+} & \alpha_{+} & \beta_{-} & \alpha_{-}  \\
                                   \alpha_{-} & \beta_{-} & \alpha_{+} & \beta_{+}\\
                                    -4+\beta_{-} &\alpha_{-} &\beta_{+} & 4+\alpha_{+}  \\
                                    \end{pmatrix},
\end{equation}
where $\alpha_{\pm}=\frac{(1-q+q\nu^{2})(n-k-1)+\nu^{2}k}{(1-q)k}\pm\frac{(1-q+\nu^{2})(k-1)}{(1-q)(n-k)}\pm\frac{q\nu^{2}}{1-q}$ and $\beta_{\pm}=\frac{(1-q-q\nu^{2})(n-k-1)+\nu^{2}k}{(1-q)k}\pm\frac{(-1+q+\nu^{2})(k-1)}{(1-q)(n-k)}\mp\frac{q\nu^{2}}{1-q}$.

If Alice measures her atom with $|pos\rangle$ at the upon arranged time $\tau_{A}=\tau$, the state of Bob's atom immediately collapses to the following form
\begin{equation}\label{Att}
  \rho^{B}_{\tau_{A}=\tau}=\gamma\begin{pmatrix}
                                             4+\alpha_{+} & \beta_{+} \\
                                            \beta_{+} & \alpha_{+}
                                           \end{pmatrix},
\end{equation}
 where $\gamma=\frac{1}{4+2\alpha_{+}}$.
When Alice's clock reads $\tau_{A}=\tau$, Bob's clock does not point at $\tau$. In other words, their local clocks have a   time difference $\delta=\tau_{A}-\tau_{B}$, which leads to the following form of Bob's state  at time $\tau_{B}=\tau$
\begin{equation}
  \rho^{B}_{\tau_{B}=\tau}=\gamma \begin{pmatrix}
                                           2+\alpha_{+}+2\cos\Omega \delta & \beta_{+}+2i\sin\Omega \delta \\
                                            \beta_{+}-2i\sin\Omega \delta&
                                             2+\alpha_{+}-2\cos\Omega \delta \\
                                           \end{pmatrix},
\end{equation}
 in the measurement basis $(|pos\rangle,|neg\rangle)$.

Therefore, the probability for Bob measuring $|pos\rangle$ at time $\tau_{B}=\tau$ is
\begin{equation}\label{probability1}
  P(|pos\rangle)=\frac{1}{2}+\frac{k(n-k)(1-q)\cos(\Omega \delta)}{(n-1)((1-q)n+\nu^{2}k+q\nu^{2}(n-k))},
\end{equation}
 which allow Bob to estimate the time difference $\delta$. We compute the  probability $P$  because: (a) The probability $P$ in Eq. (\ref{probability1}) is the only observable in our clock synchronization proposal.  (b) The time difference between Alice's standard clock and Bob's accelerated clock is included in the phase of time probability $P$. Therefore, the time difference between the accelerated clock and the standard clock is obtained if the probability is measured. (c)
As showed in Ref. \cite{Ben-Av}, the biggest  amplitude of $P$ would correspond to a more distinguished time probability and a more accurate time difference.  We  obtain the time probability $P(|neg\rangle)$ with $|neg\rangle$ to estimate the time difference $\delta$ in the same way and Bob can adjust his clock to the standard through local operation accordingly if the energy gap and time difference  satisfy the  condition  $|\Omega\delta|<2\pi$. Here we present the measurement for the time probability  $P$ with only one multipartite entangled state and for a single clock. If we want to  repeat measurement many times for quantum clock synchronization, it is necessary to provide more multipartite entangled systems, each subsystem  owned by Alice, Bob and other observers, respectively\cite{Jozsa}, and to obtain the more accurate difference of proper times between Alice and Bob.

 It is easily to know that if the number of total atoms $n=2$ and the number of excited atoms $k=1$ , the time probability in Eq.(\ref{probability1}) would also apply to two-clock systems with the initial state $|\psi^{AB}_{t_{0}}\rangle=\frac{1}{\sqrt{2}}(|01\rangle+|10\rangle)$. In other words, $Z$ state could simulate quantum clock synchronization in both two-party system and multipartite systems.
Since $W$-type state is a special kind of $Z$-type state for $k=1$,  the time probability $P_{W}(|pos\rangle)$ for $W$ state is found to be
\begin{equation}\label{probabilityW}
 P_{W}(|pos\rangle)=\frac{1}{2}+\frac{(1-q)\cos(\Omega \delta)}{((1-q)n+\nu^{2}+q\nu^{2}(n-1))}.
\end{equation}
But for $Z$-type initial state,  the parameter $k=1$  is not required to be optimized.
To obtain a higher clock adjustment accuracy, we firstly discuss the role of $k$, the number  of  excited  atoms  in the initial $Z$ state Eq. (2). After some calculations, the optimal $k$ is  found to be
\begin{equation}\label{OPTY}
  k_{opt}=\lfloor Y \rfloor
\end{equation}
for $Y=\frac{1}{(-1+q)^{2}\nu^{2}}(n(-1-q(-2+\nu^{2}))+q^{2}(-1+\nu^{2}))
+\sqrt{-n^{2}(-1+q)^{2}(-1-\nu^{2}+q^{2}(-1+\nu^{2})-q(-2+\nu^{2}))}$, where $\lfloor Y \rfloor$ takes  the nearest integer of $Y$. The optimal $k$ in Eq.(\ref{OPTY}) is required by calculating the derivative of amplitude in Eq.(\ref{probabilityW}).
Here we can see that both the coupling parameter and accelerating parameter affect the optimal choice of  $k$.

To analyze how the number of atoms and quantum  entanglement of the initial state influence the  accuracy of QCS, we distinguish the QCS in the two-body system with the two-party initial state $|\psi^{AB}_{t_{0}}\rangle$ from in the multipartite system with the $W$-type initial state or the $Z$-type initial state.
With different time probabilities for the above three initial states, we could compare variation trends of synchronization precision of quantum clock with the  different number of total atoms $n$ and the  different number of excited atoms $k$.

\begin{figure}
\begin{center}
\includegraphics[height=0.2\textheight]{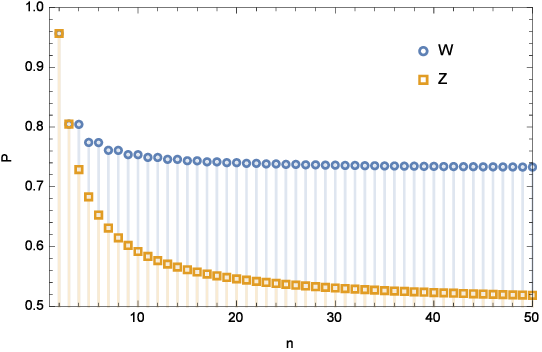}
\caption{(Color online) The time probability $P$ as a function of the number of the atom $n$ for $W$-type state and the optimal $Z$-type state with $k=\lfloor Y \rfloor$ . \iffalse A group of yellow squares with $k=1$ stand for $W$ state and a group of blue circles with $K=\lfloor Y \rfloor$ represent the optimal $Z$ state for the QCS.\fi The effective coupling parameter $\nu$ and the acceleration parameter $q$ are fixed as $\nu=0.1$ and $q=0.9$, respectively. The energy gap  $\Omega$ and the time difference $\delta$ are fixed as $\Omega\delta=2\pi$.}
\label{Fig1}
\end{center}
\end{figure}

In  Fig. 1, we analyze the time probability $P$ as a function of the number of total atoms $n$ both for  the optimal $Z$-type and $W$-type initial states.  Such  time probabilities are obtained when Bob measures  the final state of systems by $|pos\rangle$. The energy gap  $\Omega$ and the time difference  $\delta$ are fixed as $\Omega\delta=2\pi$.
In this case the  value  of time probability equals to the amplitude of probability, which is an indicator of  the clock adjustment accuracy.
It is showed that the time probability $P$, i.e., the clock adjustment accuracy of the  synchronization for $W$-type initial state equals to the accuracy  of  $Z$-type  initial state for $n=2$. In addition,  the time probability $P$  decreases  with the growth of total atoms' number $n$. This  attributes to the decrease of entanglement because, comparing to the initially bipartite fully entangled system, a pair of clocks in $Z$-type or $W$-type multipartite systems only  partly entangled, which is  the relevance of monogamy of entanglement.
 The initial entanglement has been distributed to other subsystems in multipartite systems. That is, two clocks  in an initially bipartite entangled system share more  bipartite entanglement with each other than a pair of clocks in an entangled multiparty system.  What's more, we find that the time probability $P$ of the optimal $Z$-type state with $k=\lfloor Y \rfloor$ is always more than that of $W$-type state for the same $n$. It may turn  out that the optimal $Z$ states perform better simply because the relevant entanglement is stronger in them than in $W$-type states.  In addition, it is found that the time probability $P$  approaches to different final values for different initial states in the limit of the infinite number of atoms.

To get a better understanding on how the entanglement shared between Alice and Bob influences the adjustment accuracy of the  synchronization,  we analyze the clock synchronization for a general  bipartite  entangled initial state, which has the form
\begin{equation}
|\psi^{AB}\rangle=\sin\theta|01\rangle+\cos\theta|10\rangle.
\end{equation}
Similarly, we let Alice own the static standard clock while Bob's  accelerated clock needs to be synchronized. We calculate the probability and quantum entanglement for the same state. To quantify entanglement,
we employ the concurrence \cite{Wootters,Coffman}, which is  $C(\rho)=2\max\left\{  0,\tilde{C}_{1}(\rho),\tilde{C}_{2}(\rho)\right\}$
 for a $X$-type quantum state. Here $\tilde{C}_{1}(\rho)$ and $\tilde{C}_{1}(\rho)$ are   $\tilde{C}_{1}(\rho)=\sqrt{\rho_{14}\rho_{41}}-\sqrt{\rho_{22}\rho
_{33}}$ and $\tilde{C}_{2}(\rho)=\sqrt{\rho_{23}\rho_{32}}-\sqrt{\rho
_{11}\rho_{44}}$, and $\rho_{ij}$ are  matrix elements of the  state density matrix.

The probability for the general  bipartite initial state is
\begin{equation}\label{probabilityf}
 P_{2}(|pos\rangle)=\frac{1}{2}+\frac{(1-q)\sin2\theta\cos(\Omega\delta)}{2(1-q)+2\nu^{2}(\cos^{2}\theta+q\sin^{2}\theta)}.
\end{equation}
 If we assume $\theta=\pi/4$ and the accelerated paremeter $q$ vanishes, which means the initial state is maximally entangled  and  the relativistic motion is not  considered, Eq. (\ref{probabilityf}) can be recast as $P_{2}(|pos\rangle)=\frac{1}{2}+\frac{1}{2}\cos(\Omega\delta)$, identifying with results in a non-relativistic system \cite{Jozsa,Krco,Ben-Av}.

\begin{figure}
\begin{center}
\includegraphics[height=0.2\textheight]{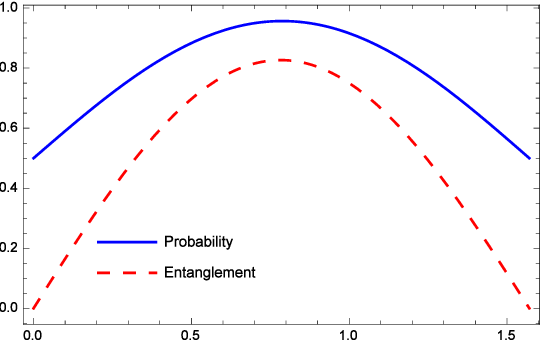}
\caption{(Color online) The measurement probability $P$ as a function of the initial state parameter $\theta$ in two-party system. The acceleration parameter  $q$ and the effective coupling parameter $\nu$ are fixed as $q=0.9$, $\nu=0.1$. The energy gap $\Omega$ and the time difference  $\delta$ are fixed as $\Omega\delta=2\pi$.}
\label{Fig2}
\end{center}
\end{figure}

In Fig. 2, we compare the time probability  and entanglement as a function of the initial state parameter $\theta$ for some fixed effective coupling parameters $\nu$. It is shown that the amplitude of time probability $P$ has similar variation trend with the entanglement as the initial state parameter $\theta$ changes. Specifically, the time probability $P$ gets the maximum when we adopt the maximal entangled initial state.
Then we conclude that the entanglement shared between the observers  enhances the accuracy of clock synchronization in the relativistic setting.

\begin{figure}
\begin{center}
\includegraphics[height=0.2\textheight]{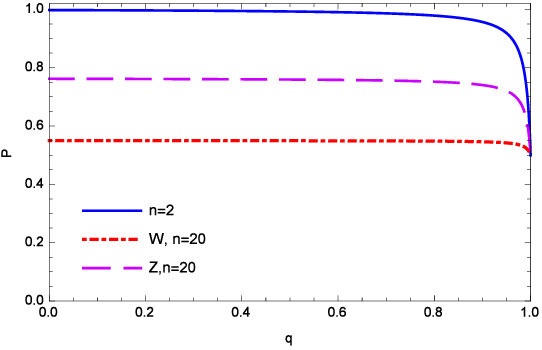}
\caption{(Color online) The time probability as a function of the acceleration parameter $q$ for bipartite state (solid line), $W$ state (dot line)  and optimal $Z$ state (dashed line), independently. The effective coupling parameter $\nu$ is fixed as $\nu=0.1$. The energy gap $\Omega$ and the time difference  $\delta$ are fixed as $\Omega\delta=2\pi$.}
\label{Fig3}
\end{center}
\end{figure}

We are also interested in how the accelerated  motion and the coupling between the atom and scalar field influence the clock adjustment accuracy.  In Fig. 3, we study the influence of acceleration parameter $q$ on time probability $P$ and compare  results of bipartite quantum clock synchronization with multipartite quantum clock synchronization (take $n=20$ as an example).  We find that the time probability $P$ sharply decreases with increasing acceleration of Bob, which reveals that the accuracy of quantum  clock synchronization is sharply reduced by  Unruh thermal noise. It is worthy note that   time probabilities for all the three different initial states approach to 0.5  when the acceleration parameter $q\rightarrow1$, i.e., in the limit of infinite acceleration. Such behavior is quite different from the dynamics of probabilities with increasing number of atoms.  Moreover, the accuracy of clock synchronization in two-party system is always better than the accuracy of multipartite systems and we find again  the $Z$-type state perform better than the $W$-type state.

\begin{figure}
\begin{center}
\includegraphics[height=0.2\textheight]{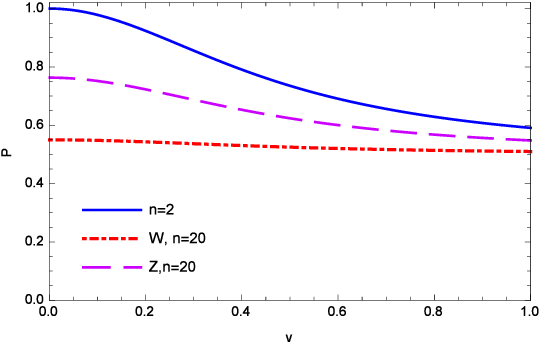}
\caption{(Color online) The time probability as a function of the effective coupling parameter $\nu$ for two-party state (solid line), $W$ state (dot line)  and optimal $Z$ state (dashed line), independently. The acceleration parameter $q$ is fixed as $q=0.8$. The energy gap $\Omega$ and the time difference  $\delta$ are fixed as $\Omega\delta=2\pi$.}
\label{Fig1}
\end{center}
\end{figure}

In Fig. 4, we show the time probability $P$ as a function of the effective coupling parameter $\nu$. We find that with increases of the effective coupling parameter $\nu$, the time probability $P$  decreases, which means that the interaction between the accelerated atom and  scalar field would reduce the clock adjustment accuracy. In addition, we find that whatever the value of the effective coupling parameter $\nu$ is, the time probability $P$ for a pair of entangled atoms is always greater than for twenty entangled atoms. Unlike the effect of acceleration, the  time probability is more robust on the coupling parameter $\nu$.

\section{conclusions}
\label{sec:conclusion}

In conclusion, we have studied the dynamic of a many Unruh-DeWitt-detector system and performed a proposal for multipartite  QCS in a relativistic setting. We have calculated  the final state of the many-detector  system  for a general $Z$-type initial state.  The information of  time difference is exposed by the observable time probability and  the clock can be adjusted accordingly.
It is shown that both the detector's acceleration and the effective coupling between Bob's accelerated detector and the massless scalar field
affect the choice of optimal number of   excited  atoms  in the initial $Z$-type state.
The clock adjustment accuracy would decrease with the increased number of the atom, which attributes to the weakness of  bipartite entanglement between Alice and Bob. Hence, the initial state with an appropriate $k$ and the less $n$ would get a higher adjustment accuracy. By comparing the behavior of time probability with that of entanglement, we find  the entanglement of  initial state would enhance the clock adjustment accuracy.  It is worthy note that   time probabilities for the optimal $Z$-type state, $W$-type state and the bipartite state we mentioned approach to the same value in the limit of infinite acceleration, while they  tend to  different minimums with increasing number of atoms. Like  Unruh thermal noise,
the interaction between the accelerated clock and the  external scalar field also reduces the accuracy of clock synchronization.

???

\begin{acknowledgments}
This work is supported by the National Natural Science Foundation
of China under Grant  No. 11675052; the Hunan Provincial Natural Science Foundation of China under Grant No. 2018JJ1016; and  the Science and Technology Planning Project of Hunan Province under Grant No. 2018RS3061. 		
\end{acknowledgments}

\end{document}